\documentclass[twocolumn,showpacs,showkeys]{revtex4}
\usepackage{graphicx}
\usepackage{bm}
\usepackage{color}
\usepackage{amsmath}
\usepackage{natbib}

\begin{document}

\title{Extraordinary SEAWs under influence of the spin-spin interaction and the quantum Bohm potential}

\author{Pavel A. Andreev}
\email{andreevpa@physics.msu.ru}
\affiliation{Faculty of physics, Lomonosov Moscow State University, Moscow, 119991, Russian Federation.}

\date{\today}

\begin{abstract}
The separate spin evolution (SSE) of electrons causes the existence of the spin-electron acoustic wave.
Extraordinary spin-electron acoustic waves (SEAWs) propagating perpendicular to the external magnetic field have large contribution of the transverse electric field.
Its spectrum has been studied in the quasi-classical limit at the consideration of the separate spin evolution.
The spin-spin interaction and the quantum Bohm potential give contribution in the spectrum extraordinary SEAW.
This contribution is studied in this paper.
Moreover, it is demonstrated that the spin-spin interaction leads to the existence of the extraordinary SEAWs if the SSE is neglected.
The hybridization of the extraordinary SEAW and the lower extraordinary wave in the regime, 
where the cyclotron frequency is larger then the Langmuir frequency is studied either.
\end{abstract}


\maketitle



\section{Introduction}

The spin-spin interaction \cite{MaksimovTMP 2001}, \cite{Hu PoP 16}, \cite{Shukla RMP 11}, the quantum Bohm potential \cite{Shukla RMP 11}, \cite{Kuzelev Ruhadze UFN1999}, \cite{Kuzelev PhUsp 11}, and the separate spin evolution (SSE) for electrons with different spin projections \cite{Andreev PRE 15 SEAW}, \cite{Andreev PRE 16}, \cite{Andreev AoP 15 SEAW} are major examples of the quantum effects in plasmas.
The extraordinary spin electron acoustic wave (SEAW) is an example of plasma wave phenomena, where all mentioned quantum effects reveal themselves.

In the linear on small amplitude of perturbations regime, the quantum Bohm potential leads to a characteristic frequency $\hbar k^{2}/2m$ which is rather small since it is proportional to the Planck constant $\hbar$.
However, it growths at the increase of the wave vector module $k$ (the short-wavelength limit).
If the wave length is comparable with the average interparticle distance $k\sim \sqrt[3]{n_{0e}}$,
where $n_{0e}$ is the equilibrium concentration of electrons,
the quantum Bohm potential is comparable with the Fermi pressure while both of them contribute in the longitudinal wave dynamics.

The spin evolution reveals in variety effects.
In simple cases the spin-spin interaction existing in the Euler equation gives a characteristic frequency which formally coincides with the characteristic frequency of the quantum Bohm potential $\hbar k^{2}/2m$.
However, the spin-spin interaction reveals in the transverse wave dynamics.
Let us mention that the spin-spin interaction is the magnetic dipole-dipole interaction \cite{Landau 2}.
In the modern science, the electrons are considered as point-like objects.
Therefore, we need to use the complete potential energy for the interaction of the magnetic dipoles corresponding to the Maxwell equations as it is presented in Ref. \cite{Landau 4}.

Moreover, the spin-spin interaction contributes in the linear wave dynamics for the magnetized plasmas while the quantum Bohm potential gives a contribution even for the unmagnetized plasmas.

Overall, the analysis of the extraordinary waves in the magnetized quantum plasmas is presented with account of the described effects.

Derivation of the models for spin evolution in quantum plasmas and first steps in studying of fundamental phenomena are performed in Refs. \cite{MaksimovTMP 2001}, \cite{MaksimovTMP 2001 b}, \cite{Oraevsky AP 02}, \cite{Oraevsky PAN}, \cite{Andreev VestnMSU 2007}, \cite{Marklund PRL07}, \cite{Brodin NJP 07}, \cite{Brodin PRL 08 Cl Reg}, \cite{Mahajan PRL 11}.
The spin evolution forms spin-plasma waves in plasmas \cite{Andreev VestnMSU 2007}, \cite{Brodin PRL 08 g Kin}, \cite{Andreev PoP 17 kin I}, \cite{Andreev PoP 17 kin II}, \cite{Andreev PoP 17 kin Non Triv} along with the SEAW \cite{Andreev AoP 15 SEAW}.
The collision spin effects in plasmas are also considered in literature (see Refs. \cite{Sasorov 15}, \cite{Sasorov 17}). 

Extraordinary SEAW can be useful in understanding of the astrophysical phenomena as it happens with other plasma waves. 
However, the SEAW is found its application in understanding of the high-temperature superconductivity 
in the magnetically ordered materials \cite{Andreev PoP 17 SupCond}. 
While other plasma physics concepts like the vorticity dynamics and the helicity conservation are useful 
in description of the superconductivity \cite{Mahajan PoP 16}.

The low frequency longitudinal waves, where the ion-motion is essential, are considered in Ref. \cite{Iqbal CTP 17}.
The spectra of the lower hybrid waves and the ion cyclotron waves obtained under the influence of the separate spin evolution are described in Ref. \cite{Iqbal CTP 17}.
Modification of the ion-acoustic solitons due to the partial spin polarization is studied in Ref. \cite{Ahmad PP 16} by means of the separate spin evolution QHD.
Influence of the partial spin polarization on the beam caused instability of the electrostatic waves is presented in Ref. \cite{Iqbal PP 17 (bdtg)}.

Quantum plasmas with the hydrodynamics and kinetics models is an example of the fields,
where the spin effects on the dynamics of electrons are studied.
The quantum field theory methods in the random phase approximation is another example where
the independent analysis of the electrons with the different spin projections is presented \cite{Ryan PRB 91}, \cite{Agarwal PRB 14}.

To the best of our knowledge, no quantum extraordinary waves caused by the spin-spin interaction has been reported to the date.
This mechanism of the extraordinary wave formation is reported in this paper.
However, it is also found that this is one of two mechanisms causing the extraordinary SEAW.
Another mechanism for the extraordinary SEAW is the SSE which is used for discovering of the extraordinary SEAW in the first place \cite{Andreev PoP 17 Extr SEAW}.

This paper is organized as follows.
In Sec. II the basic model of quantum hydrodynamics with the separate spin evolution is demonstrated.
In Sec. III the dispersion equation for the extraordinary waves is presented and the contribution of different effects is described.
In Sec. IV a numerical analysis of the dispersion equation is performed neglecting the spin-spin interaction and the quantum Bohm potential.
In Sec. V a numerical analysis of the dispersion equation is performed neglecting the SSE.
In Sec. VI a numerical analysis of the dispersion equation is presented for the general regime.
In Sec. VII a brief summary of obtained results is presented.

\section{Model}

The goal of this paper is the analysis of the spin-spin interaction and quantum Bohm potential contributions to the spectrum of the extraordinary SEAW and traditional extraordinary plasma waves.
Analysis of the SEAW requires application of the SSE-QHD \cite{Andreev PRE 15 SEAW} which is presented below.

The separate spin evolution quantum hydrodynamics contains two continuity equations
\begin{equation}\label{susdXOqBpSSI cont eq spin UP vel field} \partial_{t}n_{s}+\nabla(n_{s}\textbf{v}_{s})=(-1)^{i_{s}}\frac{\gamma_{e}}{\hbar}\varepsilon^{\alpha\beta z}S^{\alpha}B^{\beta}.  \end{equation}
One continuity equation is for the spin-up electrons and one continuity equation for the spin-down electrons. 
Interaction of electrons with the magnetic field changes the probability to find an electron in the spin-up state or the spin-down state. 
Hence, the partial numbers of electrons are changed which reveals in the nonzero right-hand side of the continuity equations.
In equation (\ref{susdXOqBpSSI cont eq spin UP vel field}) and below the following notations are used:
$n_{s}$ is the concentration of the spin-$s$ electrons, $\textbf{v}_{s}$ is the velocity field of the spin-$s$ electrons,
$S^{\alpha}$ is the spin density, $B^{\alpha}$ is the magnetic field, $i_{s}$ is 0 for the spin-up electrons and 1 for the spin-down electrons,
$\gamma_{e}$ is the magnetic moment of electron, $\hbar$ is the Planck constant, $\varepsilon^{\alpha\beta\gamma}$ is the Levi-Civita symbol,
$\partial_{t}$ is the derivative on the time variable, $\nabla$ is the gradient operator with derivatives on space variables.

There are two Euler equations in the separate spin evolution quantum hydrodynamic model:
$$mn_{s}(\partial_{t}+\textbf{v}_{s}\nabla)\textbf{v}_{s}+\nabla p_{s}
-\frac{\hbar^{2}}{4m}n_{s}\nabla\Biggl(\frac{\triangle n_{s}}{n_{s}}-\frac{(\nabla n_{s})^{2}}{2n_{s}^{2}}\Biggr)$$
\begin{equation}\label{susdXOqBpSSI Euler eq spin UP with vel} =q_{e}n_{s}\biggl(\textbf{E}+\frac{1}{c}[\textbf{v}_{s},\textbf{B}]\biggr)+\textbf{F}_{SSs},\end{equation}
with the thermal or Fermi pressure $p_{s}$, the force field of spin-spin interaction
$$\textbf{F}_{SSs}=(-1)^{i_{s}}\gamma_{e}n_{u}\nabla B_{z} +\frac{\gamma_{e}}{2}(S_{x}\nabla B_{x}+S_{y}\nabla B_{y})$$
\begin{equation}\label{susdXOqBpSSI F SS u via vel} +(-1)^{i_{s}}\frac{m\gamma_{e}}{\hbar} \varepsilon^{\beta\gamma z}\biggl(\textbf{J}_{(M)\beta}B^{\gamma}
-\textbf{v}_{s}S^{\beta}B^{\gamma}\biggr).\end{equation}
In equation (\ref{susdXOqBpSSI F SS u via vel}) and below the following notations are used:
$m$ is the mass of particle, $p_{s}$ is the partial pressure of the spin-$s$ electrons, $\triangle$ is the Laplace operator on the space variable,
$q_{e}$ is the charge of electron, $\textbf{E}$ is the electric field, $c$ is the speed of light.

In terms of the velocity field the spin current tensor is found as follows
\begin{equation}\label{susdXOqBpSSI Spin current with vel} J_{j\alpha}=\frac{1}{2}(v_{u}^{\alpha}+v_{d}^{\alpha})S_{j}
-\frac{\hbar}{4m} \varepsilon^{j\beta z}
\biggl(\frac{\partial^{\alpha} n_{u}}{n_{u}}-\frac{\partial^{\alpha} n_{d}}{n_{d}}\biggr)S_{\beta}. \end{equation}

There are two spin evolution equations for two projections of the spin density describing simultaneously all electrons disregarding the spin projections:
$$\partial_{t}S_{j}+\frac{1}{2}\nabla[S_{j}(\textbf{v}_{u}+\textbf{v}_{d})]
-\frac{\hbar}{4m}\varepsilon^{j\beta z}\nabla\Biggl(S^{\beta}\biggl(\frac{\nabla n_{u}}{n_{u}}-\frac{\nabla n_{d}}{n_{d}}\biggr)\Biggr)$$
\begin{equation}\label{susdXOqBpSSI eq for S}  +\Im_{j}
=\frac{2\gamma_{e}}{\hbar}\varepsilon^{j\beta\gamma}S^{\beta}B^{\gamma},\end{equation}
where $j=x,y$, and $\Im_{j}$ is the Fermi spin current.

The equilibrium partial Fermi pressure are used as equations of state:
\begin{equation}\label{susdXOqBpSSI Pressure spin pol} p_{s}=\frac{(6\pi^{2})^{2/3}}{5}\frac{\hbar^{2}}{m}n_{s}^{5/3}.\end{equation}

No equation of state for the Fermi spin current $\Im_{j}$ (the thermal spin current in the limit of degenerate electrons) is required in this paper
since the spin evolution does not affect the extraordinary waves.
However, equation of state for the Fermi spin current is a considerable part of the quantum hydrodynamic model.
Corresponding equation of state is considered in Refs. \cite{Andreev PoP 17 kin Non Triv}, \cite{Andreev 1510 Spin Current}, \cite{Andreev PoP 16 sep kin}.

\section{Dispersion equation for the extraordinary waves}

The extraordinary waves in the degenerate magnetized electron gas in the electron-ion plasmas are under consideration.
The constant external magnetic field is directed parallel to the Oz direction $\textbf{B}_{ext}=B_{0}\textbf{e}_{z}$.
The extraordinary waves are waves propagating perpendicular to the external magnetic field $\textbf{k}=\{k,0,0\}$ where the electric field perturbations are directed perpendicular to the external magnetic field with a components parallel to the wave vector and perpendicular to the wave vector: $\textbf{E}=\{E_{x}, E_{y}, 0\}$.

Linearized and Fourier transformed hydrodynamic equations have the following form
\begin{equation}\label{susdXOqBpSSI Cont lin} -\imath\omega\delta n_{s}+\imath kn_{0s}\delta v_{sx}=0, \end{equation}
$$-\imath\omega mn_{0s}\delta v_{sx}+\imath k\delta p_{s}+\frac{\hbar^{2}}{4m}\imath k^{3}\delta n_{s}$$
\begin{equation}\label{susdXOqBpSSI Euler lin x} =q_{e}n_{0s}(\delta E_{x}+\delta v_{sy} B_{0}/c)+(-1)^{i_{s}}\gamma_{e}n_{0s}\imath k\delta B_{z}, \end{equation}
and
\begin{equation}\label{susdXOqBpSSI Euler lin y} -\imath\omega mn_{0s}\delta v_{sy}=q_{e}n_{0s}(\delta E_{y}-\delta v_{sx} B_{0}/c). \end{equation}
So, the third projection of the linearized Euler equation gives $\delta v_{sz}=0$ since we assume $\delta E_{z}=0$ in the wave.

Equations of field give $\delta B_{z}=kc\delta E_{y}/\omega$.

Solving equations (\ref{susdXOqBpSSI Cont lin})-(\ref{susdXOqBpSSI Euler lin y}) relatively the perturbations of partial concentrations, we find the following solutions:
\begin{equation}\label{susdXOqBpSSI delta n_s} \delta n_{s}=n_{0s}\frac{\imath kq_{e}\delta E_{x}-kq_{e}\frac{\Omega_{e}}{\omega}\delta E_{y}-(-1)^{i_{s}}\gamma_{e}k^{2}\delta B_{z}}{m(\omega^{2}-\Omega_{e}^{2}-U_{Fs}^{2}-\hbar^{2} k^{4}/4m^{2})}. \end{equation}

Substitute found solutions for the concentrations and corresponding velocities in the linearized Maxwell equations which have the following form in considering regime
\begin{equation}\label{susdXOqBpSSI} -\imath\omega\delta E_{x}+4\pi q_{e}(n_{0u}\delta v_{ux}+n_{0d}\delta v_{dx})=0 \end{equation}
and
$$-\imath\omega\delta E_{y}+4\pi q_{e}(n_{0u}\delta v_{uy}+n_{0d}\delta v_{dy})$$
\begin{equation}\label{susdXOqBpSSI} -4\pi\imath kc\gamma_{e}(\delta n_{u}-\delta n_{d})=-\imath kc\delta B_{z}\end{equation}
as the result the linear evolution of small perturbations for the extraordinary waves leads to the following dispersion equation:
$$ \omega^{2}-k^{2}c^{2}-\omega_{Le}^{2}
-(\omega^{2}-\omega_{Le}^{2}-k^{2}c^{2}+\Omega_{e}^{2}+\Omega_{q\sigma}^{2})\sum_{s=u,d}\frac{\omega_{Ls}^{2}}{\Lambda_{s}}$$
\begin{equation}\label{susdXOqBpSSI Disp eq form 1} -2\Omega_{e}\Omega_{q\sigma}\sum_{s=u,d}\frac{(-1)^{i_{s}}\omega_{Ls}^{2}}{\Lambda_{s}} +\Omega_{q\sigma}^{2}\frac{\omega_{Lu}^{2}\omega_{Ld}^{2}}{\Lambda_{u}\Lambda_{d}}=0,\end{equation}
where
\begin{equation}\Lambda_{s}=\omega^{2}-\Omega_{e}^{2}-\Omega_{qB}^{2}-k^{2}U_{Fs}^{2},\end{equation}
$U_{Fs}=(6\pi^{2}n_{0s})^{1/3}\hbar/3m$ is the Fermi velocity for spin-s electrons, $\omega_{Le}^{2}=4\pi e^{2}n_{0e}/m$ is the square of the Langmuir frequency, $\Omega_{e}=eB_{0}/mc$ is the cyclotron frequency, $n_{0e}=n_{0u}+n_{0d}$.
Quantum frequencies $\Omega_{qB}$ and $\Omega_{q\sigma}$ are equal to each other $\Omega_{qB}=\Omega_{q\sigma}=\hbar k^{2}/2m$, but they have different sources.
Frequency $\Omega_{qB}$ appears from the quantum Bohm potential while frequency $\Omega_{q\sigma}$ comes from the spin-spin interaction force field.

Dispersion equation (\ref{susdXOqBpSSI Disp eq form 1}) can be represented in the following form
$$ \omega^{2}-k^{2}c^{2}
- \sum_{s=u,d}\frac{\omega_{Ls}^{2}}{\Lambda_{s}}(2\omega^{2}-\omega_{Le}^{2}-k^{2}c^{2}-k^{2}U_{Fs}^{2})$$
\begin{equation}\label{susdXOqBpSSI Disp eq form 2} -2\Omega_{e}\Omega_{q\sigma}\sum_{s=u,d}\frac{(-1)^{i_{s}}\omega_{Ls}^{2}}{\Lambda_{s}} +\Omega_{q\sigma}^{2}\frac{\omega_{Lu}^{2}\omega_{Ld}^{2}}{\Lambda_{u}\Lambda_{d}}=0,\end{equation}
where the third and fourth terms of equation (\ref{susdXOqBpSSI Disp eq form 1}) are combined together and relation $\Omega_{qB}=\Omega_{q\sigma}$ is used.

\section{Regime of the SSE without the spin-spin interaction and the quantum Bohm potential}

\begin{figure}
\includegraphics[width=8cm,angle=0]{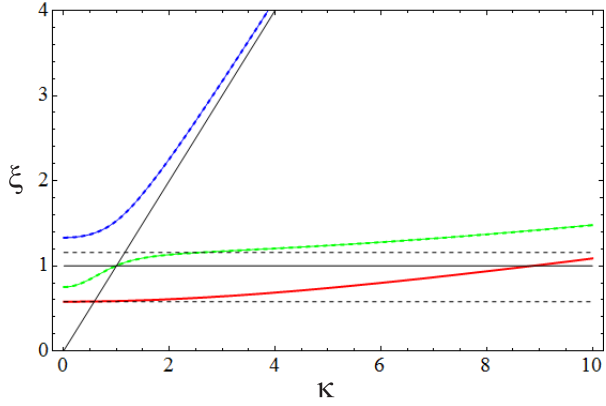}
\caption{\label{susdXOqBpSSI_01} Figure shows the spectrum of extraordinary waves in quantum plasmas without the spin-spin interaction and the quantum Bohm potential at $\mid\Omega_{e}\mid<\omega_{Le}$.
Three bold continues curves describe three extraordinary waves existing in quantum plasma due to the SSE account.
Two upper curves are the traditional extraordinary waves.
The third extraordinary wave is the extraordinary SEAW presented by the lower (red) curve.
The upper (blue) and middle (green) dashed curves describe the traditional extraordinary waves with no account of the SSE.
There are four asymptotic lines:
The horizontal dashed lines show $\xi=0.58$, and $\xi=1.16$ which corresponds to $\omega=\mid\Omega_{e}\mid$, and $\omega=\sqrt{\omega_{Le}^{2}+\Omega_{e}^{2}}$;
The thin horizontal continuous line shows $\xi=1$ which corresponds to $\omega=\omega_{Le}$;
The inclined continuous line shows $\xi=\kappa$ which corresponds to $\omega=kc$.
This figure is plotted for $n_{0e}=2.6\times10^{27}$ cm$^{-3}$, $B_{0}=10^{11}$ G, $\eta_{e}=0.09$.}
\end{figure}

\begin{figure}
\includegraphics[width=8cm,angle=0]{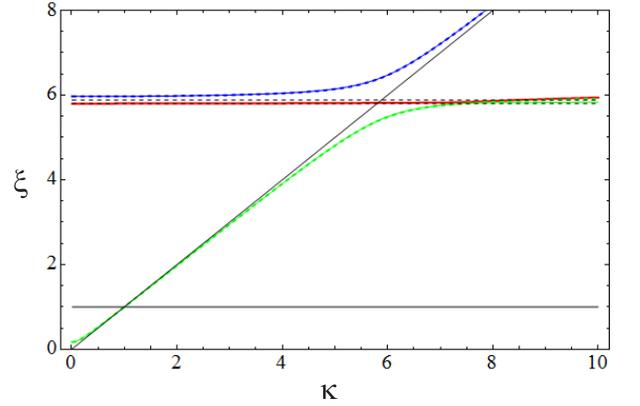}
\caption{\label{susdXOqBpSSI_02} Figure shows the spectrum of extraordinary waves in quantum plasmas without the spin-spin interaction and the quantum Bohm potential at $\mid\Omega_{e}\mid>\omega_{Le}$.
Three bold continues curves describe three extraordinary waves existing in quantum plasma due to the SSE account.
The upper (blue) and the lower (green) curves are the traditional extraordinary waves.
The third extraordinary wave is the extraordinary SEAW presented by the middle (red) curve.
The upper (blue) and the lower (green) dashed curves describe the traditional extraordinary waves with no account of the SSE.
There are four asymptotic lines:
The horizontal dashed lines show $\xi=5.8$, and $\xi=5.9$ which corresponds to $\omega=\mid\Omega_{e}\mid$, and $\omega=\sqrt{\omega_{Le}^{2}+\Omega_{e}^{2}}$;
The thin horizontal continuous line shows $\xi=1$ which corresponds to $\omega=\omega_{Le}$;
The inclined continuous line shows $\xi=\kappa$ which corresponds to $\omega=kc$.
This figure is plotted for $n_{0e}=2.6\times10^{27}$ cm$^{-3}$, $B_{0}=10^{12}$ G, $\eta_{e}=0.72$.}
\end{figure}

Regime of the SSE without the spin-spin interaction and the quantum Bohm potential is considered in Ref. \cite{Andreev PoP 17 Extr SEAW}.
In this regime the dispersion equation (\ref{susdXOqBpSSI Disp eq form 2}) simplifies to
$$\omega^{2}-k^{2}c^{2} -\sum_{s=u,d}\frac{\omega_{Ls}^{2}}{\omega^{2}-\Omega_{e}^{2}-k^{2}U_{Fs}^{2}}$$
\begin{equation}\label{susdXOqBpSSI Disp eq form 2 no qF} \times(2\omega^{2}-\omega_{Le}^{2}-k^{2}c^{2}-k^{2}U_{Fs}^{2})=0. \end{equation}
Equation (\ref{susdXOqBpSSI Disp eq form 2 no qF}) is the equation of the third degree relatively $\omega^{2}$.
If there is no difference between the spin-up electrons and the spin-down electrons equation (\ref{susdXOqBpSSI Disp eq form 2 no qF}) simplifies to the traditional equation of the second degree relatively $\omega^{2}$.
It demonstrates that the SSE leads to additional wave: the extraordinary SEAW \cite{Andreev PoP 17 Extr SEAW}.
Corresponding numerical analysis is demonstrated in Figs. \ref{susdXOqBpSSI_01} and \ref{susdXOqBpSSI_02} for two different regimes $\mid\Omega_{e}\mid<\omega_{Le}$ and $\mid\Omega_{e}\mid>\omega_{Le}$ correspondingly.
The fact that the SSE causes an extra wave at the propagation of waves perpendicular to the external field even in the electrostatic regime was discovered earlier \cite{Andreev AoP 15 SEAW}.
However, the transverse part of the electric field gives a considerable contribution and changes the dispersion dependence \cite{Andreev PoP 17 Extr SEAW}.

Coincidence of the dashed blue curve (dashed green curve) with the continuous blue curve (continuous green curve) demonstrated in Figs. \ref{susdXOqBpSSI_01}
shows small contribution of the SSE in the traditional extraordinary waves.
The same correct for the dashed blue curve in Fig. \ref{susdXOqBpSSI_02} (the upper extraordinary wave).
The lower extraordinary wave in the second regime $\mid\Omega_{e}\mid>\omega_{Le}$ has a hybridization with the SEAW.

The extraordinary SEAW caused by the SSE is presented by the red curves in Figs. \ref{susdXOqBpSSI_01}
and \ref{susdXOqBpSSI_02} (the lower continuous curve in Fig. \ref{susdXOqBpSSI_01} or the middle continuous curve in Fig. \ref{susdXOqBpSSI_02}).

It has been shown in Ref. \cite{Andreev PoP 17 Extr SEAW} that the frequency of the extraordinary SEAW is considerably larger then the frequency of corresponding electrostatic SEAW.
Moreover, the phase velocity of the extraordinary SEAW is considerably larger
then the phase velocity of the electrostatic SEAW either.

\subsection{Electrostatic SEAW with the quantum Bohm potential}

\begin{figure}
\includegraphics[width=8cm,angle=0]{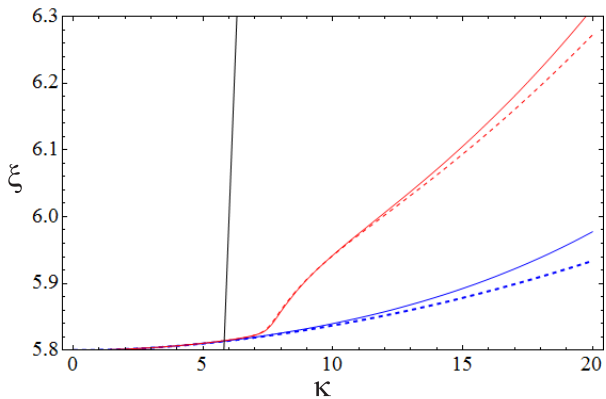}
\caption{\label{susdXOqBpSSI_03}
The figure presents a comparison of the electrostatic SEAW and the extraordinary SEAW.
Dashed curves do not include the quantum Bohm potential while the continuous curves include the quantum Bohm potential.
The lower (the upper) dashed curve describes the electrostatic (the extraordinary) SEAW.
The lower (the upper) continuous curve presents the electrostatic (the extraordinary) SEAW.
This figure is plotted for $n_{0e}=2.6\times10^{27}$ cm$^{-3}$, $B_{0}=10^{12}$ G, $\eta_{e}=0.72$.}
\end{figure}

Electrostatic limit can be directly found from the general dispersion equation (\ref{susdXOqBpSSI Disp eq form 2}) assuming
that the interaction happens with infinite speed $c\rightarrow\infty$.
Hence, equation (\ref{susdXOqBpSSI Disp eq form 2}) simplifies to the following equation
\begin{equation}\label{susdXOqBpSSI Disp eq form 1 electrostatic} 1-\sum_{s=u,d}\frac{\omega_{Ls}^{2}}{\Lambda_{s}}=0.\end{equation}
This equation gives the electrostatic hybrid wave
which has cut-off at $\omega(k=0)=\sqrt{\omega_{Le}^{2}+\Omega_{e}^{2}}$.
Moreover, this equation has the second solution which is the SEAW.
The electrostatic SEAW propagating perpendicular to the external magnetic field is considered in Ref. \cite{Andreev AoP 15 SEAW} with no account of the quantum Bohm potential.
For the complete picture, it is necessary to compare the spectrum of the electrostatic SEAW and the spectrum of the extraordinary SEAW under influence of the quantum Bohm potential.
This comparison is presented in Fig. (\ref{susdXOqBpSSI_03}).
As it is expected from general properties of the quantum Bohm potential it gives contribution in the short-wavelength limit.
The quantum Bohm potential contributes in both cases: the electrostatic SEAW and the extraordinary SEAW.
In both cases it increases frequency of the waves.
However, the contribution of the transverse part of the electric field
(transition from the blue curves to the red curves) found in Ref. \cite{Andreev PoP 17 Extr SEAW} is considerably larger.
Hence, the comparison of solutions of equations (\ref{susdXOqBpSSI Disp eq form 2 no qF})
and (\ref{susdXOqBpSSI Disp eq form 1 electrostatic}) is presented in Fig. (\ref{susdXOqBpSSI_03}).

\section{Regime of the spin-spin interaction and the quantum Bohm potential without the SSE}

Let us consider the regime,
where the spin-spin interaction and the quantum Bohm potential are included,
but the separate spin evolution is disregarded.
It means that all electrons are considered as the single fluid. 
However, a trace of the SSE QHD application exists here. 
So, this approximation can be called quasi-single fluid model (see discussion in Appendix A).
For this regime the general dispersion equation (\ref{susdXOqBpSSI Disp eq form 2}) can be simplified
by the assumption of the zero spin polarization
(it corresponds to $\omega_{Lu}^{2}=\omega_{Ld}^{2}=\omega_{Le}^{2}/2$, $U_{Fu}^{2}=U_{Fd}^{2}=U_{Fe}^{2}$).
It leads to the next equation:
$$\omega^{2}-k^{2}c^{2}+\Omega_{q\sigma}^{2}\frac{\omega_{Le}^{4}}{\Lambda_{e}^{2}}$$
\begin{equation}\label{susdXOqBpSSI Disp eq form 2 no SSE}
-\frac{\omega_{Le}^{2}}{\Lambda_{e}}(2\omega^{2}-\omega_{Le}^{2}-k^{2}c^{2}-k^{2}U_{Fe}^{2})=0,\end{equation}
where
\begin{equation}\Lambda_{e}=\omega^{2}-\Omega_{e}^{2}-\Omega_{qB}^{2}-k^{2}U_{Fe}^{2}.\end{equation}

If we neglect the spin-spin interaction (means to drop $\Omega_{q\sigma}$)
we loose the third term in equation (\ref{susdXOqBpSSI Disp eq form 2 no SSE}).
It reveals in the decreasing of the degree of the dispersion equation.
Hence, the spin-spin interaction is a cause of the extra wave appearance.

If the quantum Bohm potential is neglected
(means to drop $\Omega_{qB}$) the small wavelength behavior of three wave solutions is changed,
but all solutions exist.

Due to the nature of the extra wave appearance let us call it
the spin-dipole extraordinary wave.

\begin{figure}
\includegraphics[width=8cm,angle=0]{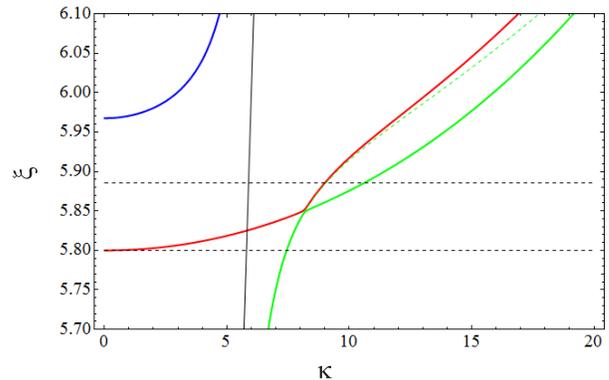}
\caption{\label{susdXOqBpSSI_04} Figure shows the spectrum of extraordinary waves in quantum plasmas in the regime,
where the quantum Bohm potential and the spin-spin interaction are included
in addition to the classical effects while all electrons are considered as the single fluid. 
Three bold continues curves describe three extraordinary waves existing in quantum plasma due to the quantum Bohm potential and the spin-spin interaction account.
The upper (blue) and the lower (green) curves are the traditional extraordinary waves.
The third extraordinary wave is the extra extraordinary wave caused by the spin-spin interaction
and presented by the middle (red) curve.
The upper (blue) and the lower (green) dashed curves describe the traditional extraordinary waves with no account of the quantum Bohm potential and the spin-spin interaction (bold and dashed blue lines coincide).
There are three asymptotic lines:
The horizontal dashed lines show $\xi=5.8$, and $\xi=5.9$ which corresponds to $\omega=\mid\Omega_{e}\mid$, and $\omega=\sqrt{\omega_{Le}^{2}+\Omega_{e}^{2}}$;
The inclined continuous line shows $\xi=\kappa$ which corresponds to $\omega=kc$.
This figure is plotted for $n_{0e}=2.6\times10^{27}$ cm$^{-3}$, $B_{0}=10^{12}$ G, $\eta_{e}=0.72$.}
\end{figure}

\begin{figure}
\includegraphics[width=8cm,angle=0]{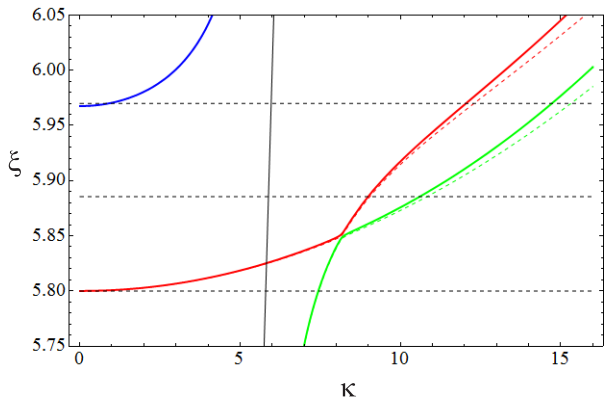}
\caption{\label{susdXOqBpSSI_05}
Figure shows the spectrum of extraordinary waves in quantum plasmas in the regime where the quantum Bohm potential is included
in addition to the classical effects and the spin-spin interaction
while all electrons are considered as the single fluid. 
Three bold continues curves describe three extraordinary waves existing in quantum plasma due to the spin-spin interaction account
under influence of the quantum Bohm potential.
The upper (blue) and the lower (green) curves are the traditional extraordinary waves.
The third extraordinary wave is the extra extraordinary wave caused by the spin-spin interaction
and presented by the middle (red) curve.
The upper (blue) and the lower (red) dashed curves describe the extraordinary waves
(the upper traditional extraordinary wave and the extra extraordinary wave caused by the spin-spin interaction, correspondingly)
with no account of the quantum Bohm potential (and no account of the SSE).
There are four asymptotic lines:
The horizontal dashed lines show $\xi=5.8$, $\xi=5.88$, and $\xi=5.97$ which corresponds to $\omega=\mid\Omega_{e}\mid$,
$\omega=\sqrt{\omega_{Le}^{2}+\Omega_{e}^{2}}$ and $\omega=\sqrt{2\omega_{Le}^{2}+\Omega_{e}^{2}}$.
The third asymptotic line $\xi=\kappa$ coincides with the green continuous inclined line describing lower extraordinary wave.
Here $\omega=\sqrt{2\omega_{Le}^{2}+\Omega_{e}^{2}}$ is the large cyclotron frequency asymptotic value for the upper extraordinary wave frequency
at the zero wave vector $\omega^{2}_{UE}(0)=0.5(2\omega_{Le}^{2}+\Omega_{e}^{2}+\sqrt{4\omega_{Le}^{2}\Omega_{e}^{2}+\Omega_{e}^{4}})$.
This figure is plotted for $n_{0e}=2.6\times10^{27}$ cm$^{-3}$, $B_{0}=10^{12}$ G, $\eta_{e}=0.72$.}
\end{figure}

Modification of spectrum caused by the simultaneous contribution of the spin-spin interaction
and the quantum Bohm potential in compare with the classic spectrum is presented in Fig. \ref{susdXOqBpSSI_04}.
Moreover, the influence of the quantum Bohm potential is considered in addition to the spin-spin interaction in Fig. \ref{susdXOqBpSSI_05}.


Coincidence of the upper continuous curve and the upper dashed curve in Fig. \ref{susdXOqBpSSI_04} shows that
the combined action of the spin-spin interaction and the quantum Bohm potential gives no noticeable changes of the frequency of the upper extraordinary wave.
As in the classic regime the upper extraordinary wave frequency goes asymptotically to $kc$ from the area of larger frequencies at the large wave vectors.
The lower extraordinary wave frequency in the classic regime is presented by the lower dashed curve in Fig. \ref{susdXOqBpSSI_04}.
It goes just below the middle (red) continuous curve.
The middle (red) continuous curve in Fig. \ref{susdXOqBpSSI_04} presents the spin-dipole extraordinary wave spectrum.
Fig. \ref{susdXOqBpSSI_04} demonstrates the hybridization of the spin-dipole extraordinary wave and the lower extraordinary wave
which happens at $\kappa\approx8.4$.
At $\kappa>8.4$ the spin-dipole extraordinary wave is similar to the classic lower extraordinary wave.
However, the middle (red) continuous curve goes higher then the lower dashed curve
due to simultaneous action of the spin-spin interaction and the quantum Bohm potential.
The frequency of the lower extraordinary wave at $\kappa>8.4$ is decreases in compare with the classical regime
due to the hybridization with the spin-dipole extraordinary wave.

The spin-spin interaction is included in all curves in Fig. \ref{susdXOqBpSSI_05} as an addition to the classical effects.
There is no change in the upper extraordinary wave spectrum.
The green (single) dashed line in Fig. \ref{susdXOqBpSSI_04} coincides with the red (upper) dashed line in Fig. \ref{susdXOqBpSSI_05}.
The hybridized lower extraordinary wave frequency at $\kappa>8.4$ is increased by the quantum Bohm potential
as it follows from comparison of the lower (green) dashed curve and the lower (green) continuous curve.

If there is no account of the transverse part of the electric field the extra wave caused by the spin-spin interaction disappears.

\section{General regime for the quantum extraordinary waves}

In this section the general form of the dispersion equation (\ref{susdXOqBpSSI Disp eq form 2}) is analyzed.

First of all, it is necessary to notice that the SSE or the spin-spin interaction leads to formation of an extra wave.
In the first case the wave is called the extraordinary SEAW.
In the second case it is called the spin-dipole extraordinary wave.
However, the general dispersion equation (\ref{susdXOqBpSSI Disp eq form 2}) has same degree as equations (\ref{susdXOqBpSSI Disp eq form 2 no qF}) and (\ref{susdXOqBpSSI Disp eq form 2 no SSE}).
Hence, in general case,
the extraordinary SEAW and the spin-dipole extraordinary wave are the same wave existing
due to the composition of two effects.
We keep calling this wave the extraordinary SEAW since
the extraordinary SEAW exists in the electrostatic limit while the spin-dipole extraordinary wave does not have any trace in the electrostatic limit
(it exist in the narrower interval).

Figs. (\ref{susdXOqBpSSI_01}) and (\ref{susdXOqBpSSI_02}) show small influence of the SSE on the traditional upper extraordinary wave.
On the other hand,
the lower extraordinary wave at $\mid\Omega_{e}\mid>\omega_{Le}$ is affected by the SSE due to the hybridization with the SEAW.
The hybridization of the lower extraordinary wave and the SEAW (the spin-dipole extraordinary wave) happens due to the spin-spin interaction in the absence of the SSE.

Therefore, it is necessary to consider the simultaneous action of the SSE, the spin-spin interaction,
and the quantum Bohm potential on the lower extraordinary wave and the SEAW.

\begin{figure}
\includegraphics[width=8cm,angle=0]{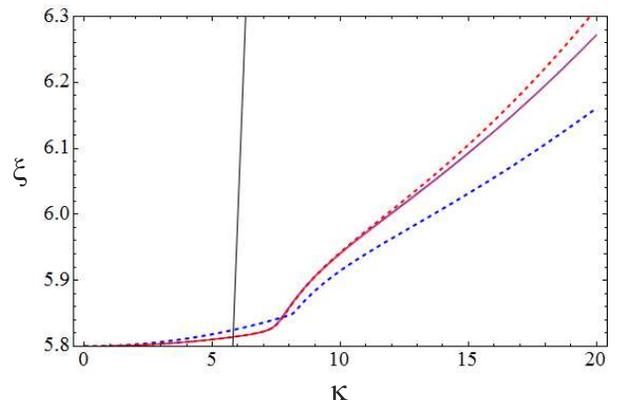}
\caption{\label{susdXOqBpSSI_06}
Figure shows details of the spectrum the extraordinary SEAW.
The spectrum is demonstrated in four regimes.
The continuous (blue) curve describes the extraordinary SEAW caused by the SSE.
The continuous (red) curve shows the extraordinary SEAW under simultaneous influence of the SSE and the spin-spin interaction (the two-fluid regime).
The lower dashed (blue) curve presents the extraordinary SEAW
(the spin-dipole extraordinary wave) caused by the spin-spin interaction in the single fluid regime.
The upper dashed (red) curve describes the extraordinary SEAW under simultaneous influence of the SSE,
the spin-spin interaction, and the quantum Bohm potential (the two-fluid regime).
The inclined (almost vertical) continuous line shows $\xi=\kappa$ which corresponds to $\omega=kc$.
This figure is plotted for $n_{0e}=2.6\times10^{27}$ cm$^{-3}$, $B_{0}=10^{12}$ G, $\eta_{e}=0.72$.}
\end{figure}

\begin{figure}
\includegraphics[width=8cm,angle=0]{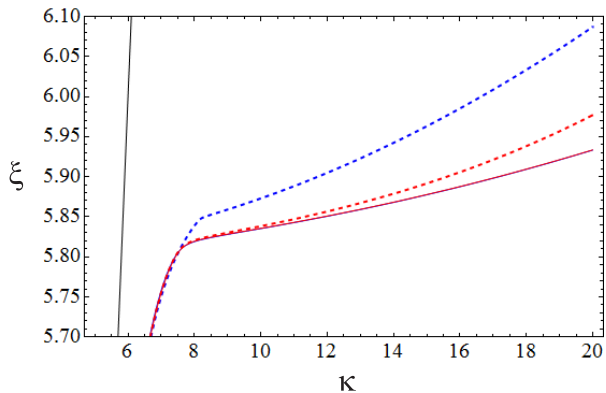}
\caption{\label{susdXOqBpSSI_07}
Figure shows details of the spectrum the lower extraordinary wave (under hybridization).
The spectrum is demonstrated in four regimes.
The continuous (blue) curve describes the extraordinary SEAW caused by the SSE.
The continuous (red) curve shows the extraordinary SEAW under simultaneous influence of the SSE and the spin-spin interaction (the two-fluid regime).
The continuous curves almost coincide to each other.
The upper dashed (blue) curve presents the extraordinary SEAW
(the spin-dipole extraordinary wave) caused by the spin-spin interaction in the single fluid regime.
The lower dashed (red) curve describes the extraordinary SEAW under simultaneous influence of the SSE,
the spin-spin interaction, and the quantum Bohm potential (the two-fluid regime).
The inclined (almost vertical) continuous line shows $\xi=\kappa$ which corresponds to $\omega=kc$.
This figure is plotted for $n_{0e}=2.6\times10^{27}$ cm$^{-3}$, $B_{0}=10^{12}$ G, $\eta_{e}=0.72$.}
\end{figure}

Influence of various effects on these waves is demonstrated in Figs. \ref{susdXOqBpSSI_06} and \ref{susdXOqBpSSI_07} 
and described in the captions to these figures.

\begin{figure}
\includegraphics[width=8cm,angle=0]{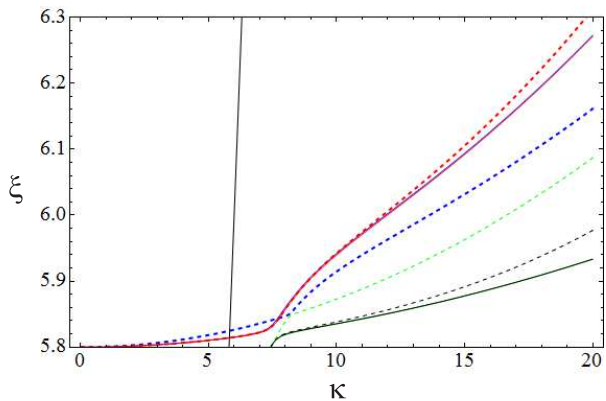}
\caption{\label{susdXOqBpSSI_08}
Figure shows details of the spectra of the extraordinary SEAW and the lower extraordinary wave (under hybridization).
The spectrum is demonstrated in four regimes and presents the combination of Figs. \ref{susdXOqBpSSI_06} and \ref{susdXOqBpSSI_07}.}
\end{figure}

Consider the lower dashed (blue) curve in Fig. \ref{susdXOqBpSSI_06}, the upper dashed (blue) curve in Fig. \ref{susdXOqBpSSI_07},
and corresponding curves in Fig. \ref{susdXOqBpSSI_08} which present the extraordinary SEAW
(the spin-dipole extraordinary wave) caused by the spin-spin interaction in the single fluid regime.
We find that difference between two curves
(in the area of hybridization, where they are getting close to each other) is rather small in compare with other regimes,
where the SSE is included.
The wave vectors and frequencies of the converging point are changed (decreased) under influence of the SSE.

\section{Conclusion}

The separate spin evolution leads to the spin-electron acoustic wave.
If waves propagate parallel to the external field the SEAW has electrostatic nature.
Analysis of electrostatic SEAW propagating perpendicular to the external magnetic field demonstrates its existence either.
Account of the transverse part of the electric field demonstrates existence of the extraordinary SEAW.
The spectrum of the extraordinary SEAW is considerably differs from the spectrum of the electrostatic SEAW.
The extraordinary SEAW exists along with the traditional extraordinary waves.

Influence of the spin-spin interaction and the quantum Bohm potential on
the traditional extraordinary waves and the extraordinary SEAW has been considered.

While the quantum Bohm potential gives a contribution in the short-wavelength part of the extraordinary SEAW.
The spin-spin interaction and the separate spin evolution have key role in the appearance of the extraordinary SEAW.

The numerical analysis is performed for the relatively dense degenerate plasmas located in the high magnetic field.
Hence, the presented results have been obtained for the dense astrophysical objects like the white dwarfs and neutron stars.
However, the major conclusions about the role of the SSE, the spin-spin interaction, and the quantum Bohm potential
are relevant for various degenerate plasmas.
In particular, it can be applied to the magnetically ordered metals and semiconductors,
where the conductivity electrons and holes can have large spin polarization in the absence of the large external magnetic field.

\section{Appendix A: Relation between the single fluid model and the quasi-single fluid model}

The spin-spin interaction force in linearized form has the following form in the single fluid regime $\textbf{F}_{SSI}=\imath \textbf{k}\gamma_{e}S_{0}\delta B_{z}$, where $S_{0}$ is the z-projection of the equilibrium spin density. 
Parameter $S_{0}$ can be represented via the ratio of the spin polarization and the concentration: $S_{0}=\eta n_{0}$. 
Therefore, the contribution of the spin-spin interaction in the concentration perturbation $\delta n_{e}$ 
and the velocity field perturbations $\delta v_{ex}$ and $\delta v_{ey}$ is proportional to $S_{0}\sim\eta$. 
However, the sum/difference of $\delta n_{u}$ and $\delta n_{d}$ from equation (\ref{susdXOqBpSSI delta n_s}) 
(obtained for the two-fluid model of electrons) shows a different picture.

The last term in the numerator of $\delta n_{u}+\delta n_{d}$ is proportional to $n_{0u}-n_{0d}\sim\eta$. 
While, the last term in the numerator of $\delta n_{u}-\delta n_{d}\sim\delta M_{z}$ in the Maxwell equations is proportional to $n_{0u}+n_{0d}=n_{0e}$, 
where no spin polarization is appeared. 
This feature creates difference between the single fluid model and the quasi-single fluid model which is reduction of the two-fluid model of electrons.

\end{document}